\documentclass[a4paper,prl,twocolumn,superscriptaddress, reprint]{revtex4-1}
\pdfoutput=1

\usepackage[colorlinks,linkcolor={blue},citecolor={blue},urlcolor={blue}]{hyperref}

\usepackage{graphicx}

\usepackage{bm,amsmath}

\usepackage{latexsym}

\usepackage{verbatim}

\usepackage{color}

\usepackage{subfigure}

\usepackage{gensymb}

\newcommand{\rb}{{\bf r}}

\newcommand{\eb}{{\bf e}}

\newcommand{\Fb}{{\bf F}}
\newcommand{\Ib}{{\bf I}}
\newcommand{\mb}{{\bf m}}
\newcommand{\Tb}{{\bf T}}

\newcommand{\vb}{{\bf v}}

\newcommand{\Omegab}{\mbox{\boldmath $\Omega$\unboldmath}}
\newcommand{\omegab}{\mbox{\boldmath $\omega$\unboldmath}}

\newcommand{\mutens}{\mbox{\boldmath $\mu$\unboldmath}}

\newcommand{\beq}{\begin{equation}}

\newcommand{\eeq}{\end{equation}}

\newcommand{\bea}{\begin{eqnarray}}

\newcommand{\eea}{\end{eqnarray}}

\newcommand{\hs}[1]{\textcolor{blue}{#1}}

\begin{document}

\title{Zipping and Entanglement in Flagellar Bundle of \emph{E.\ Coli}: Role of Motile Cell Body}

\author{Tapan Chandra Adhyapak}
\email{\tt tapan.c.adhyapak@tu-berlin.de}
\affiliation{Institut f\"{u}r Theoretische Physik, Technische Universit\"{a}t
Berlin, Hardenbergstrasse 36, 10623 Berlin, Germany}
\author{Holger Stark}
\email{\tt holger.stark@tu-berlin.de}
\affiliation{Institut f\"{u}r Theoretische Physik, Technische Universit\"{a}t
Berlin, Hardenbergstrasse 36, 10623 Berlin, Germany}

\date{\today}

\begin{abstract}

The course of a peritrichous bacterium such as \emph{E. coli} crucially depends
on the level of synchronization and self-organization of several rotating
flagella. However, the rotation of each flagellum generates counter body
movements which in turn affect the flagellar dynamics. Using a detailed
numerical model of an \emph{E. coli}, we demonstrate that flagellar
entanglement, besides fluid flow relative to the moving body, dramatically
changes the dynamics of flagella from that  compared to anchored flagella.  In
particular, bundle formation occurs through a zipping motion in a remarkably
rapid time, affected little by initial flagellar orientation. A simplified
analytical model supports our observations.  Finally, we illustrate how
entanglement, hydrodynamic interactions, and body movement contribute to
zipping and bundling.

\end{abstract}


\maketitle

Understanding self-propulsion of microorganisms pose utmost challenges
involving rich and complex physics \hs{\cite{raima, luga2009,
coretz,saragosti2011, rodenborn2013, strong, davod2015}.} Bacteria are among
the simplest and widely studied of such systems \cite{turner2000, turner2007,
macnab1977, macnab2003, rhodobacter1, rhodobacter2}. Yet, only recently we are
able to explore in full detail the underlying physics involved
\cite{larson2015, jonathan2015, philipp2014, reinhardPRL2013,  gerhardPLOS2013,
gomperSM2012, reinhardEPJE2012, grahamPRE2011, larson2010, reinhardPB,
wada2008, coretz2005, reichartEPJE2005}; much of this, however, is still to be
apprehended.  Propulsion of peritrichous bacteria such as \emph{E.\ coli} is
generated by the rotation of a bundle of several helical propellers, called
flagella. Flagella are passive filaments rotated at one end by rotary motors
embedded in the cell wall \cite{berg_ecoli}.  The level of synchronization and
self-organization of rotating flagella crucially decides the swimming course of
the cell body, to which they are attached, making it either propel or tumble.
However, the dependency is not one-sided: the rotary motors that rotate each
flagellum also produce systematic body movements, which in turn affect the
flagellar dynamics.  While the rotating cell body drags the proximal ends of
flagella with it, the distal ends cannot keep up due to friction with the
surrounding fluid.  In understanding flagellar synchronization and bundling
dynamics, focus has so far been given primarily on hydrodynamic interactions
and elastic properties of flagella \cite{powers2003, powersPRE2004, coretz2005,
reichartEPJE2005, coretz2005, grahamPRE2011, gomperSM2012}. Although, body
movement is speculated to play an important role too \cite{powersPRE2002,
gomperSM2012}, knowledge of its detailed impact is still lacking.   

The cell body moves in response to the forces and torques acting on it
\cite{purcell1997}. It translates due to the thrust force generated by the
rotating bundle of flagella in the surrounding fluid medium at low Reynolds
number.  It also has to rotate since the torque on the body has to balance all
motor torques acting on the flagella.  Ref.\ \cite{powersPRE2002} argued that
the sole effect of body rotation on a flagellum is to simply wrap it around the
cell-body axis and thereby enhance bundling.  However, real flagella are
helical, during tumbling they are more or less arbitrarily oriented
\cite{turner2000}, and they cannot simply pass through each other.  We will
demonstrate through our simulations that body rotation in such situations leads
to entanglement, where portions of different flagella obstruct each other's
free course.  While it is well established that hydrodynamic interactions are
sufficient to synchronize \cite{reichartEPJE2005} and bundle
\cite{grahamPRE2011, gomperSM2012, gerhardPLOS2013} anchored flagella, the role
of entanglement for flagellar dynamics is not known so far.  In particular, an
understanding how cell body motion influences flagellar dynamics is incomplete
without considering flagellar entanglement.

In this paper we explore how body movements influences the dynamics of flagella
with the help of a realistic numerical model of an \emph{E. coli} that includes
detailed flagellar elasticity, hydrodynamic and steric interactions among
flagella, and a motile cell body. We demonstrate that body movements
dramatically change flagellar behavior leading to profound impacts on the
overall dynamics of the cell.  In particular, bundle formation happens through
flagellar synchronization and a `zipping' motion on an experimental time scale
\cite{turner2000}, which is much smaller than for anchored flagella
\cite{grahamPRE2011,gomperSM2012} and which is approximately independent of the
initial orientations of unbundled flagella.  We also analyze the relative
importance of body movements and flagellar interactions for synchronization and
bundling.  Our work therefore is a major step towards understanding the
propulsion of a peritrichous bacterium close to its real conditions
\cite{turner2000, turner2007}.

We first summarize our approach to describe the dynamics of the cell body with
multiple flagella in an unbounded fluid of viscosity $\eta$.  We treat each
flagellum as a slender body with centerline $\rb(s)$ parametrized by the arc
length $s$.  By affixing the orthonormal tripod $\left\{\eb_1(s), \eb_2(s),
\eb_3(s)\right\}$ at each point on the centerline, where $\eb_3$ is the local
tangent and $\eb_1$ and $\eb_2$ are unit vectors along the principal axes of
the flagellar cross section, one can fully characterize the bent and twisted
flagellum.  Dynamics of the flagellum now is governed by Langevin equations for
$\rb(s)$ and the twist angle $\phi(s)$ about the centerline
\cite{reinhardPRL2013}:
\begin{eqnarray}
\partial_t \rb &=& \mutens_t \left(\Fb_{\mathrm{el}} + \Fb_{\mathrm{s}} + 
                   \Fb_{\mathrm{th}}\right) + \vb_{\mathrm{h}} , \label{r_eqn} \\
\partial_t \phi &=& \mu_r (T_{\mathrm{el}} + T_{\mathrm{th}}), \label{phi_eqn}  
\end{eqnarray}
Here, we separate the velocity contribution $\vb_{\mathrm{h}}$ due to
hydrodynamic interactions from local terms and denote local forces and torques
by $\Fb$'s and $T$'s, respectively.  Self-mobilities $\mutens_t = \eb_3 \otimes
\eb_3/\gamma_\parallel + (\Ib - \eb_3\otimes\eb_3)/\gamma_\perp$ and $\mu_r =
1/\gamma_R$ are expressed in terms of friction coefficients per unit length.
For the flagellum of an \emph{E.\ coli} they are $\gamma_\parallel = 1.6 \times
10^{-3} \mathrm{pNs}/\mu \mathrm{m}^2$, $\gamma_\perp = 2.8 \times 10^{-3}
\mathrm{pNs}/\mu \mathrm{m}^2$ and $\gamma_R = 1.26 \times 10^{-6}
\mathrm{pNs}$ \cite{reinhardEPJE2012}.  Thermal forces $\Fb_{\mathrm{th}}$ and
torques $T_{\mathrm{th}}$ are shown for completeness. Although these are
predicted to play an important role during rotation-induced polymorphic
transformations of a flagellum \cite{reinhardEPJE2012}, they are negligible in
our present study \cite{grahamPRE2011} and are ignored.   

\begin{figure}
\includegraphics[width=0.7\columnwidth]{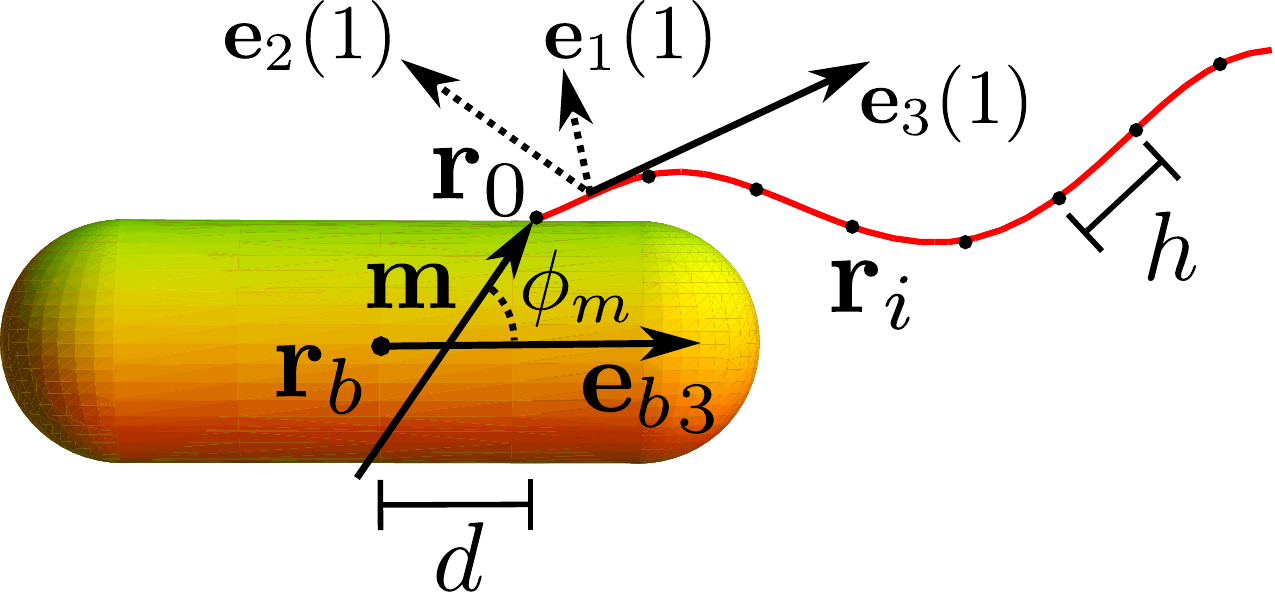}

\caption{Schematics of the cell body and part of one flagellum. For clarity,
length $h$ between discretization points $r_i$ is out of proportion and only
the first material tripod  $\{\eb_1(1), \eb_2(1), \eb_3(1)\}$ is shown.  The
cell body is centered at $\rb_b$ and oriented along the vector ${\eb_b}_3$.
The flagellum is attached at $\rb_{0}$ off-centered by a distance $d$ along
${\eb_b}_3$.  The motor torque drives a tripod at $\rb_{0}$.}

\label{schematic_model}
\end{figure}

Elastic forces and torques $ \Fb_{\mathrm{el}} = -\delta \mathcal{F}/\delta
\rb$ and $T_{\mathrm{el}} = -\delta \mathcal{F}/\delta \phi$ are derived,
respectively, from the total elastic free energy $ \mathcal{F} \left[\rb(s),
\phi(s) \right]$ of the flagellum, the form of which is obtained as follows.
The rotational strain vector $\Omegab$ moves the material tripod along the
flagellum: $\partial_s \eb_{\nu} = \Omegab \times \eb_{\nu};\,\nu=1,2,3$.
Therefore, its components completely characterize the instantaneous
flagellar conformation \cite{reinhardEPJE2012}.  A small deformation $d\Omegab
= \Omegab - \Omegab^0$ from the normal helical ground state $\Omegab^0 =
\left\{0.0, 1.3, -2.1\right\}$ \cite{reinhardPRL2013, calladine1976} needs the
Kirchhoff elastic free energy density \cite{landau_elasticity},
$f_{\mathrm{k}}(\Omegab) = (A/2) \left[ (d\Omega_1)^2 + (d\Omega_2)^2 \right] +
(C/2) (d\Omega_3)^2 $.  For \emph{E. coli}, we choose an isotropic bending
rigidity$ A=3.5 \, \mathrm{pN} \mu \mathrm{m}^2$ (assuming a circular flagellar
cross section) and the twist rigidity $C = A$ \cite{turner2007}. Integrating
over the length of the flagellum, we obtain
$\mathcal{F}\left[\rb(s),\phi(s)\right]= \int ds \left(f_{\mathrm{k}} +
f_{\mathrm{st}} \right)$, where we include a stretching free energy density
$f_{\mathrm{st}} = K(\partial_s \rb)^2/2$ with $K = 10^3 \mathrm{pN}$
\cite{reinhardPRL2013}. 

To proceed, we discretize Eqs.\ (\ref{r_eqn}) and (\ref{phi_eqn}) by
considering discrete positions $\rb_i\equiv\rb(s_i)$ along each flagellum and
by assigning $\left\{\eb_1(i), \eb_2(i), \eb_3(i)\right\}$ to the straight
segment of length $h$ between $\rb_{i-1}$ and $\rb_{i}$ (see Fig.\
\ref{schematic_model}).  Excluded-volume interactions among flagella are
enforced by the steric force $\Fb_s(\rb_i) = \sum_j \Fb_s^j (h-h_j)/h$.  Here,
the summation runs over all overlaps occurring within $[\rb_{i-1},\rb_{i+1}]$
of a given flagellum and $\Fb_s^j$ is the steric force at a distance $h_j < h$
from $\rb_i$, appropriately decomposed to act on the discrete points (for
details see the supplemental material \cite{supp1}). $\Fb_s^j$ derives from the
Lennard-Jones potential $U_{LJ}(r_m) = (F_0 \sigma/6) \, [\left(\sigma/r_m
\right)^{12} - \left( \sigma/ r_m \right)^{6}]\, \Theta(2^{1/6}\sigma - r_m)$.
We truncate it at the mimimum using the heaviside step function $\Theta(x)$,
where $r_m$ is the minimal distance between the approaching two flagellar
centerlines and $F_0$ is the strength of the steric force at $r_m = \sigma$.
We choose $F_0 = 0.8$ pN and adjust $\sigma = 4 a$, with $a$ the
cross-sectional radius of the flagellar filament, ensuring numerical stability
during entanglement in all situations.  

Finally, to include hydrodynamic interactions between the flagella, we treat
each discrete point $\rb_i$ as a sphere of radius $a$ and set $\vb_{\mathrm{h}}
(\rb_i) = \sum_{j\ne i} \mutens_{ij} \Fb(\rb_j)$.  Here, $\mutens_{ij}$ is the
Rotne-Prager mobility matrix \cite{dhont} for spheres at $\rb_i$ and $\rb_j$,
$\Fb(\rb_j)$ is the local force at $\rb_j$, and the summation runs over all
points of both flagella. We neglect subleading effects from hydrodynamic
interactions due to rotation of the spheres.  Furthermore, neglecting
hydrodynamic lubrication for close flagella is justified for thin filaments and
the presence of asperities in real flagella \cite{lubri}.

We model the cell body by a spherocylinder of length $L_b=2.5\,\mu\mathrm{m}$
and width $d_b = 0.8\,\mu\mathrm{m}$ \cite{berg_ecoli} (see Fig.\
\ref{schematic_model}).  Point $\rb_{0}$ of each flagellum is fixed on the body
surface and a motor torque $\Tb_m = T_m \mb$ drives the flagellum by rotating
the motor tripod $\left\{\eb_1(0), \eb_2(0), \eb_3(0)= \mb \right\}$ at
$\rb_{0}$.  This tripod couples to the main part of the flagellum through the
Kirchhoff elastic free energy density $f_{\mathrm{k}}$, where we set $A\to0$
and $C\to3 C$.  Thus, the driving torque is transferred to the flagellum
through a `hook' that acts as an universal joint with low bending and high
twist rigidities \cite{hook_universaljoint} allowing the first flagellar
segment along $\eb_3(1)$ to be at any angle to $\mb$.  In response, the body
moves and rotates with velocities $\vb_b = \mutens_b^t \Fb_b$ and $\omegab_b =
\mutens_b^r (\Tb_b + \Tb_m)$, respectively.  Here, $ \Fb_b$ and $\Tb_b$ are the
respective force and torque (relative to the body center) resulting from the
forces $\Fb_{\mathrm{el}} + \Fb_{\mathrm{s}}$ that act on the flagellar
anchoring points.  For the mobilities $\mutens_b^t$ and $\mutens_b^r$ we use
the analytically available values for a prolate spheroid of aspect ratio
$L_b/d_b$ \cite{kim_karilla}. The angle $\phi_m$ between $\mb$ and ${\eb_b}_3$
is expected to differ from $90\degree$ because of, for example, a locally
curved body surface.  We adjust $\phi_m = 55\degree$ to obtain a ratio for the
bundle-to-body rotation within the experimentally observed range
\cite{turner2007}. Furthermore, we employ the same potential $U_{LJ}(r_m)$ to
describe the excluded volume interaction between the body and flagella, where
$r_m$ now is the minimal distance between the body surface and any $\bf{r}_i$
on a flagellum.

\begin{figure} 

\includegraphics[width=0.85\columnwidth]{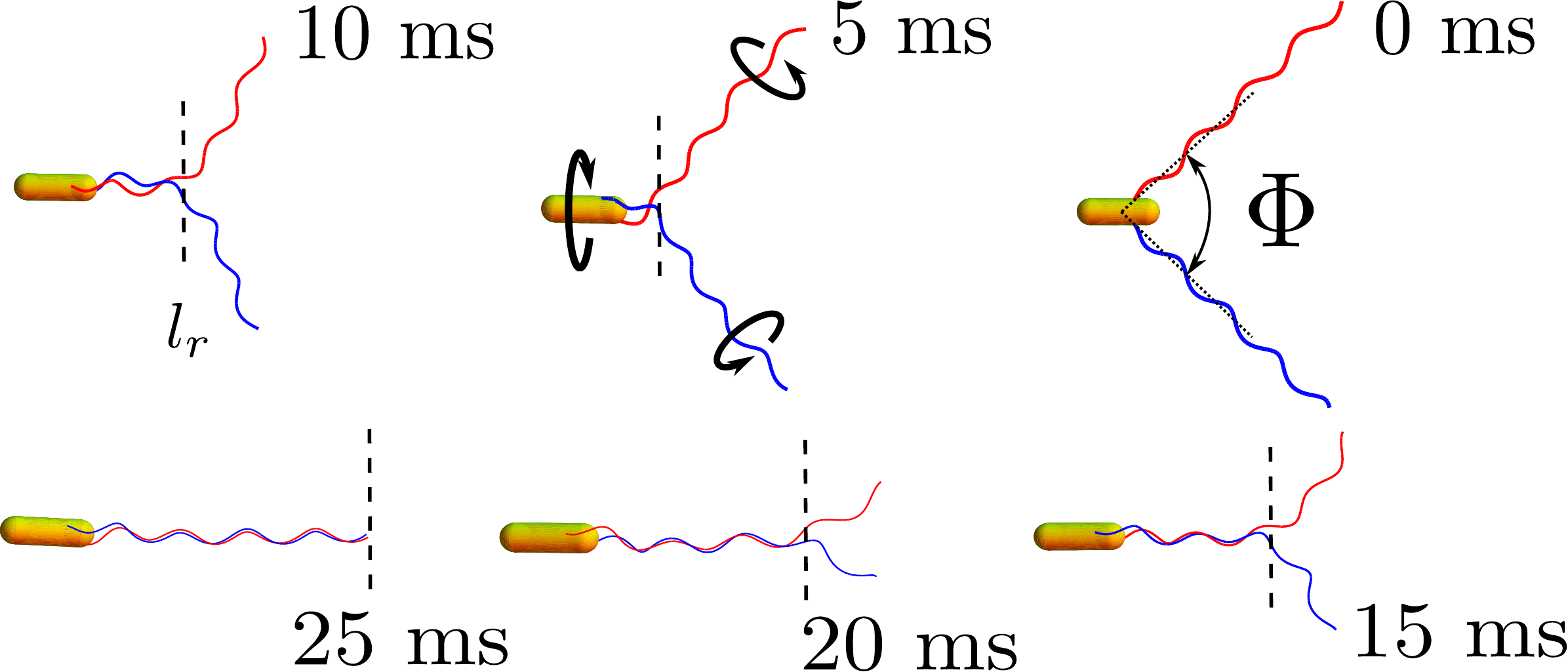}

\caption{ As the bacterium moves to the left, the flagella form a bundle of
length $l_r$ similar to  a closing zipper. Snapshots at various times $t$ (in
ms) for $\Phi=90\degree$ are shown. Rotation directions of cell body and
flagella are depicted.}

\label{snapshots} 
\end{figure}
 
\begin{figure}
\includegraphics[width=0.95\columnwidth]{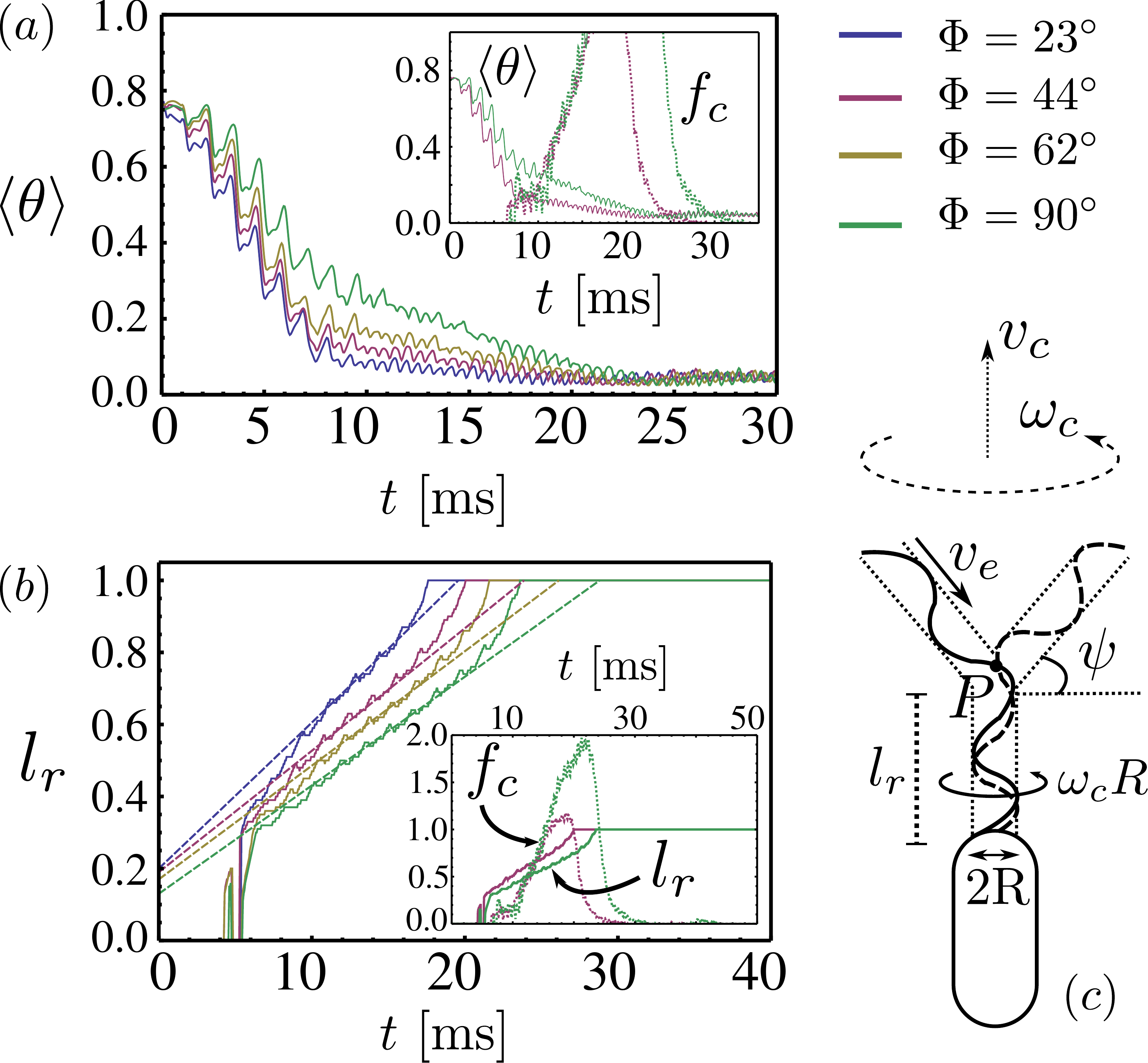}

\caption{(a) Contour-averaged phase difference $\langle\theta\rangle$ plotted
versus time $t$ for various values of the opening angle $\Phi$.  (b) Bundle
length $l_r$ plotted against $t$ for the same $\Phi'$s obtained from
simulations (full lines) and an analytical model (dashed lines). The analytical
model fits best for $R=0.30\,\mu\mathrm{m}$, when $v_c =
22.5\,\mu\mathrm{m}/s$, $\omega_c/2\pi= 11.4$ Hz, $\lambda_f =
2.5\,\mu\mathrm{m}$, $\nu_f= 83.3$ Hz. (c) Body motion ($v_c$, $\omega_c$) and
flagellar phase velocity $v_e$ determine the speed of the bundle front $P$ in
the analytic model.  Insets: Steric force density $f_c$ (in $\mathrm{pN/\mu
m}$) correlated with time evolutions of $\langle\theta\rangle$ in (a) and $l_r$
in (b), for $\Phi = 44\degree$ and $90\degree$.  }

\label{observations}
\end{figure}

We now turn to our observations.  Fig.\ \ref{snapshots} shows typical snapshots
of the bacterium moving towards the left with the flagella in their normal
left-handed helical form.  The snapshots were obtained at regular intervals
from a particular simulation run.  Quantitative details of the corresponding
flagellar dynamics are presented in Fig.\ \ref{observations}.  At time $t= 0$
ms, flagella start with an angle $\Phi$ between their axes.  As time
progresses, they rotate counterclockwise about their axes (as viewed from
behind the cell) when driven by a positive motor torque $T_m = 3.4 $ pN$\mu$m
\cite{turner2007, reinhardPRL2013}.  Simultaneously, the cell body performs a
counterbalancing clockwise rotation and also translates because of the thrust
force generated by the flagella.  The resultant flagellar evolution is complex.
It involves entanglement and large bending of flagellar axes. To quantify
synchronization of flagellar rotation in such situations, we need to compare
the respective tripod vectors $\eb_{\nu}(i)$ and $\tilde{\eb}_{\nu}(i)$ from
the two flagella at the same flagellar position $i$. Therefore, we introduce
the effective phase difference $\theta(i) = \cos^{-1} \left[ \eb_1(i) \cdot
\tilde{\eb}_1^*(i) \right]/\pi$ with $\tilde{\eb}_1^*(i)=
\mathcal{R}[\tilde{\eb}_1(i)]$, where $\mathcal{R}$ rotates the tangent vectors
onto each other [$\tilde{\eb}_3^*(i) = \eb_3(i)$] about the axis $\eb_3(i)
\times \tilde{\eb}_3(i)$.  Starting from a non-zero initial value, the contour
average $\langle \theta \rangle$ quickly drops towards zero and the flagella
reach a nearly synchronized rotational state after about $10 \,
\mathrm{ms}$ [see Fig.\ \ref{observations}(a)].  This is also reflected in
the snapshot at $10 \, \mathrm{ms}$ shown in Fig.\ \ref{snapshots}, where the
phases of both flagella clearly match.  The initial regime remains unaffected
when changing the initial value of $\langle \theta \rangle$.  It is
completely determined by hydrodynamics since steric forces are zero as
documented by the inset of Fig.\ \ref{observations}(a).  The subsequent, almost
linear decrease of $\langle \theta \rangle$ towards full synchronization
coincides with bundle formation, which we discuss now.

While cell-body movements only quantitatively change synchronization dynamics,
they dramatically influence bundling dynamics compared to the case of anchored
flagella \cite{grahamPRE2011, gomperSM2012, gerhardPLOS2013}.  At $t \sim
10\,\mathrm{ms}$ proximal portions of the flagella start to wrap around the
body axis [see Fig. \ref{snapshots}], after synchronization has already
proceeded considerably.  This marks the beginning of bundling near the cell
body, while rest of the flagella are still apart. With time the front of the
bundled portion advances away from the body, gradually  drawing remaining loose
flagellar portions into the bundle.  The growth of the bundled portion thus
resembles a `zipping' motion, where the `zip' starts at a point near the body
and continues till whole of the flagella have joined the bundle. 

To quantify these findings, we define the bundled portion as the part of a
flagellum for which all its points are at distances $\le 2 R$ from the other
flagellum.  We choose $R= 0.22\,\mu\mathrm{m}$ to be the equilibrium radius of
the helical flagellum.  The bundle length $l_r$ is then measured from the cell
body to the bundle front normalized by the axial length $L_{ax}$ of a
flagellum.  As seen from Fig.\ \ref{observations}(b), `zipping' takes place
within approximately 5 - $20 \, \mathrm{ms}$.  During this period, after an
initial sharp increase $l_r$ grows almost linearly with time.  Furthermore, for
the studied range of opening angles $\Phi$ the bundling time only varies by
about $10 \, \mathrm{ms}$.  Therefore, it is always much smaller than the total
tumble time of 150 - $450\,\mathrm{ms}$ \cite{turner2000}.  So, bundling during
a tumble event is remarkably independent of the extent to which a flagellum is
thrown out of a bundle.  This gives a crucial insight into the locomotion of a
bacterium since bundling without supporting body movements takes much longer.

Flagellar entanglement is observed to play an important role in all the
findings mentioned above. During zipping, the steric force density $f_c =
|\Fb_s|/L$, where $L$ is the length of a flagellum, is found to build up until
bundling completes, after which $f_c$ declines rapidly [inset, Fig.\
\ref{observations}(b)].  Due to viscous drag and flagellar flexibility, parts
of the flagella not in the bundle hardly follow the cell-body rotation (see
movie M1 in the Supplemental material \cite{supp1}).  As a result, proximal
portions of the flagella not only start to wrap around the body axis but also
get entangled [snapshot at $t=10 \, \mathrm{ms}$, Fig.\ \ref{snapshots}].  This
drastically enhances flagellar bundling: further rotation of proximal ends is
possible only when the entangled front proceeds away from the body, gradually
bringing the rest of the flagella quickly into the bundle.  A simplified
analytical model discussed below further supports these observations.  A
correlation between $f_c$ and the slow linear decline of $\langle \theta
\rangle$ mentioned earlier is clear from the inset of Fig.\
\ref{observations}(a).  While flagella in the bundle are synchronized, the
local phase differences $\theta(i)$ of their free ends fluctuate strongly until
entanglement forces them into the bundle.  This is observed to be responsible
for the delayed slow decline of $ \langle \theta \rangle $.

A simplified model for the zipping dynamics takes into account flagellar
entanglement at the bundle front and fluid flow in the body-fixed reference
frame. The latter occurs with respective translational and rotational
velocities $\vb_c = - \vb_b$ and $\omegab_c= - \omegab_b$ [see Fig.\
\ref{observations}(c)].  Accordingly, the length of the bundle grows with speed
$d l_r/dt$ and the angle $\psi = 90\degree - \Phi/2$ varies in time according
to
\begin{equation} 
\frac{d l_r}{dt}  = v_{\mathrm{e}} + \frac{\omega_c R}{\cos{\psi}} 
\quad\enspace \mathrm{and} \quad\enspace
\frac{d \psi}{dt} = \frac{v_c\cos{{\psi}}}{L_{ax} - l_r}  \, .
\label{eq.l_r_psi}
\end{equation}
Like in a zipper the free portions of the rotating flagella are dragged into
the bundle front $P$ with the helical phase velocity $v_{\mathrm{e}}$. They are
perfectly fit into the bundle; hence $l_r$ grows with the same speed
$v_{\mathrm{e}} = \lambda_f \nu_f$, where $\lambda_f$ is the helical pitch and
$\nu_f$ the frequency of rotation. Second, the surrounding fluid wraps the free
flagellar portions with an angular velocity $\omega_c$ onto the bundle
cylinder. So, flagella are dragged with an additional speed $\omega_c
R/\cos{\psi}$ into the bundle front, where $R$ is an effective bundle radius.
Finally, the translational flow rotates the flagellar tip with speed $v_c
\cos{\psi}$ about the bundle front $P$, which gives the angular velocity $d\psi
/ dt$.  Eqs.\ (\ref{eq.l_r_psi}) are solved for $l_r(t)$ using parameter values
measured from our simulations with $R$ and $l_r(t=0)$ adjusted for the best
fit. The results plotted as dashed lines in Fig.\ \ref{observations}(b) agree
well with our simulation results, strengthening the interpretation of our
observations discussed above.

\begin{figure} \includegraphics[width=8cm]{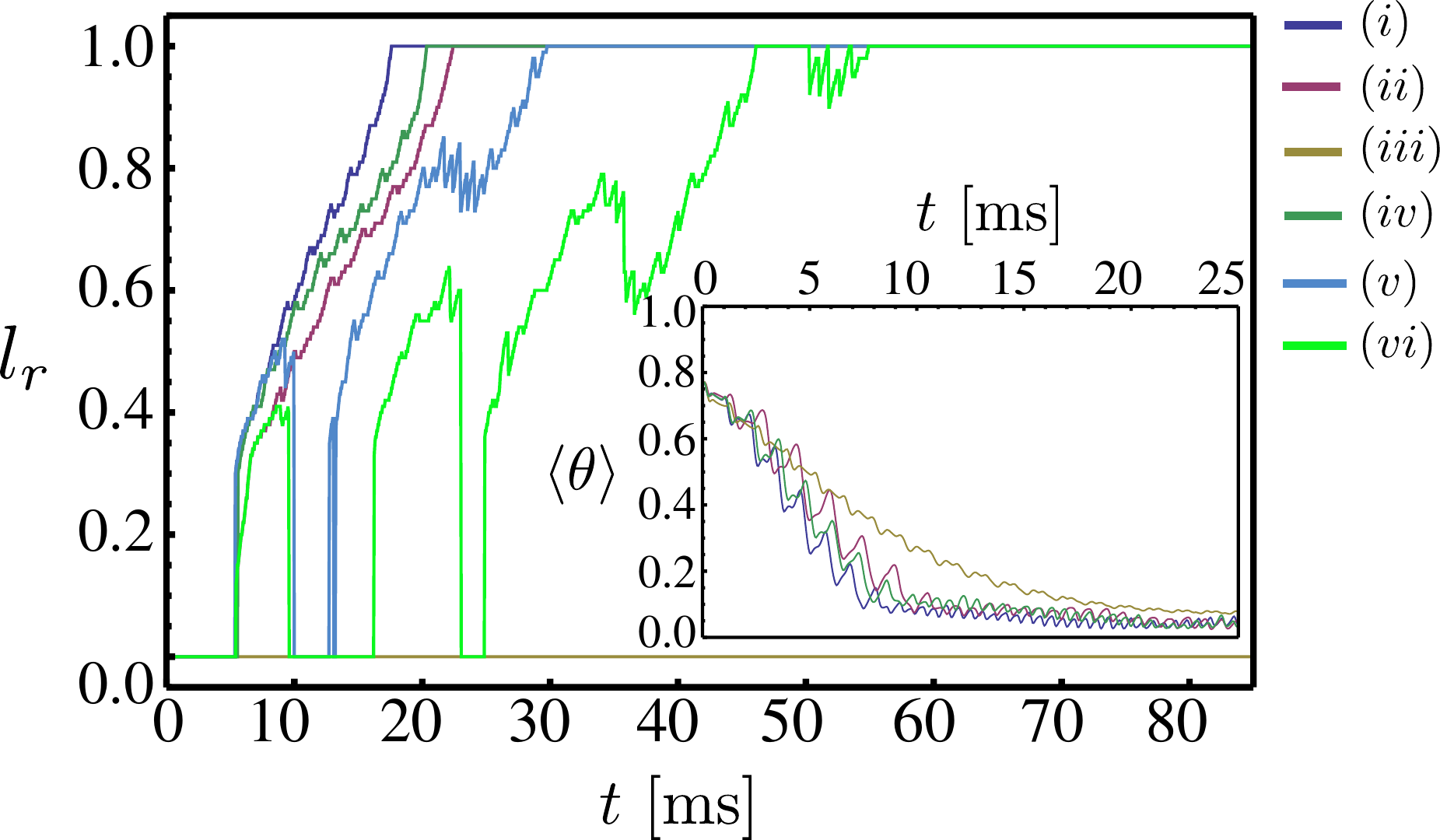}

\caption{Comparison of different factors affecting bundling and synchronization
dynamics (inset).  Curves (i): standard simulation run performed for $\Phi=23
\degree$, $\theta(i) = 0.8 \, \forall i$, and $T_m = 3.4\,\mathrm{pN\mu m}$.
Other curves show simulation runs that ignore: (ii) hydrodynamic interactions
among flagella, (iii) axial body rotation, (iv) body translation, (v) steric
interaction between flagella, or (vi) both steric and hydrodynamic
interactions.}

\label{impacts}
\end{figure}

To obtain further insight into flagellar dynamics, we make a comparative study
to judge the relevance of cell-body movements and various flagellar
interactions.  The results are presented in Fig.\ \ref{impacts}.  We consider
the standard simulation run performed at $\Phi=23 \degree$.  Its results
[curves (i) in Fig.\ \ref{impacts}] are compared with those obtained from new
simulations where either body movements or flagellar interactions are ignored.
The impact of body rotation on the synchronization dynamics is more pronounced
than that of body translation [inset, curves (iii) and (iv)].  Without body
rotation, $\langle \theta \rangle$ converges more slowly towards zero,
qualitatively resembling the results of anchored flagella
\cite{reichartEPJE2005}.  However, absence of body translation does not affect
the outcome much.  Hydrodynamic interactions (HI) between flagella are
important in reproducing the standard result of the full simulation [inset,
curves (i) and (ii)].  However, we find that synchronization is even possible
without HI contrasting the situation of anchored flagella, where HI between
flagella are known to be essential \cite{reichartEPJE2005}. Interestingly,
cell-body movement is sufficient to synchronize flagella, similar to
findings for \emph{Chlamydomonas} in Ref.\ \cite{friedrichPRL2012}.

Bundling dynamics is affected with a similar trend. While bundling is delayed
by ca.\ $10 \, \mathrm{ms}$ in absence of HI between flagella, absence of body
translation is less severe [curves (i), (ii), and (iv) in Fig.\ \ref{impacts}].
There is no bundle formation in absence of axial body rotation [ horizontal
line as curve (iii)] because the body re-orients and slows down translation,
resulting in an unusual buckling of flagella away from each other.  More
significantly, when we follow Ref.  \cite{powersPRE2002} and allow temporal
evolution of each flagella affected only by body movement but not by either HI
or steric interactions, bundling gets significantly delayed by about
$30\,\mathrm{ms}$ [curves (v) and (vi)].     

To conclude, we present a detailed modeling of an \emph{E. coli} with two
flagella and a motile cell body.  This allows us to probe bacterial propulsion
on an yet experimentally inaccessible level. The complex role of flagellar
polymorphism is ignored for simplicity.  In principle, it can be probed
extending our model \cite{reinhardPRL2013}.  We demonstrate that compared to
the situation of anchored flagella, flagellar dynamics close to real conditions
is strikingly altered by body movements.  Times to bundle and synchronize are
dramatically reduced.  In particular, flagellar entanglement helps bundling to
proceed quickly like a `zipping' motion, which we rationalize in a simplified
model.  Furthermore, we demonstrate that body movement and flagellar
entanglement lead to rapid bundling and synchronization even when hydrodynamic
interactions are neglected.  Our findings are important in explaining
experimentally observed times scales as
mentioned
above.

Finally, more and more artificial microswimmers using different swimming
mechanisms have been and are constructed \cite{artificial1, *artificial2,
*artificial3}.  We provide here an example how one develops a model for
exploring  and ultimately understanding the biomechanics of microswimmers.

\acknowledgments

We are grateful to R. Vogel for useful discussions and providing key insights
to the numerics involved.  We also acknowledge D. Alizadehrad, G. Gompper, P.
Kanehl, O. Pohl, C. Prohm, R. Winkler, and A.  Z\"ottl for helpful discussions.
We thank the VW foundation for financial support within the program
``Computational Soft Matter and Biophysics" (Grant No. I/86 801). 

%
\end{document}